\documentclass[prb,showpacs]{revtex4}
\usepackage{amsmath}
\usepackage{latexsym}
\usepackage{float}
\usepackage{amssymb}
\usepackage{graphicx}
\usepackage{epsfig}

\begin{document}

\newcommand{\be}{\begin{equation}}
\newcommand{\ee}{\end{equation}}
\newcommand{\ba}{\begin{eqnarray}}
\newcommand{\ea}{\end{eqnarray}}

\title{Universal properties of a single polymer chain in slit: Scaling versus MD
simulations}

\author{D.I.Dimitrov,}\affiliation{Inorganic Chemistry and Physical
Chemistry Department,University of Food Technology,Maritza Blvd.26,4002
Plovdiv,Bulgaia}
\author{A.Milchev,}\affiliation{Institute for Chemical Physics,Bulgain Academy
of Sciences,1113 Sofia  Bulgaia and Institut f\"ur Physik, Johannes
Gutenberg-Universit\"at Mainz\\ D-55099 Mainz, Staudinger Weg 7, Germany}
\author{Kurt Binder}
\affiliation{Institut f\"ur Physik, Johannes Gutenberg-Universit\"at Mainz\\
D-55099 Mainz, Staudinger Weg 7, Germany}
\author{Leonid I. Klushin}
\affiliation{American University of Beirut, Department of Physics,
Beirut, Lebanon}
\author{Alexander M. Skvortsov}
\affiliation{Chemical-Pharmaceutical Academy, Prof. Popova 14, 197022
St. Petersburg, Russia.}

\date{\today}

\begin{abstract}

We revisit the classical problem of a polymer confined in a slit in both of its
static and dynamic aspects. We confirm a number of well known scaling
predictions and analyse their range of validity by means of comprehensive
Molecular Dynamics simulations using a coarse-grained bead-spring model of a
flexible polymer chain.

The normal and parallel components of the average end-to-end distance, mean
radius of gyration and their distributions, the density profile, the force
exerted on the slit walls, and the local bond orientation characteristics  are
obtained in slits of width $D$ = $4 \div 10$ (in units of the bead radius)
and for chain lengths $N=50 \div 300$. We demonstrate that a wide range of
static chain properties in normal direction can be described {\em
quantitatively} by analytic model - independent expressions in perfect agreement
with computer experiment. In particular, the observed profile of
confinement-induced bond orientation, is shown to closely match theory
predictions.

The anisotropy of confinement is found to be manifested most dramatically in
the dynamic behavior of the polymer chain. We examine the relation between
characteristic times for translational diffusion and lateral relaxation. It is
demonstrated that the scaling predictions for lateral and normal relaxation
times are in good agreement with our observations. A novel feature is the
observed coupling of normal and lateral modes with two vastly different
relaxation times.

We show that the impact of grafting on lateral relaxation is equivalent to
doubling the chain length.

\end{abstract}

\pacs {36.20-r, 36.20.Ey, 02.70.Lq}

\maketitle

\section{Introduction}
One of the most impressive successes in the theory of polymer solutions was the
discovery of a close analogy with critical phenomena in ferromagnetic
systems~\cite{deGennes,deGennes2,Freed,Cloizeaux,Schaefer}: scaling theory based
on this analogy predicts that in the case of a good solvent and flexible
polymers the variation with molecular mass and concentration of many properties
directly measured experimentally can be presented in universal form as functions
of a dimensionless parameter - the ratio of a characteristic length to the
average size of a polymer coil (Flory radius~\cite{Flory}) $R_F=aN^\nu$ with the
Flory index $\nu\approx 3/5$ in dilute solution. Here $a$ is the size of an
effective monomeric unit and $N$ - the number of such units in a polymer chain
whereby, for simplicity, a prefactor of an order of unity is suppressed. For
example, the mean-squared distance between monomers, mean-squared radius of
gyration, static structure factors,  were presented in scaling
form~\cite{Cloizeaux}. The results were also generalized to include the effects
of varying spatial dimensionality, solvent quality, and chain stiffness. Most of
these predictions were verified both experimentally and by computer simulations.
In fact, for an athermal isolated flexible macromolecule in three or two
dimensions the level of description is so detailed that it includes not only the
average characteristics but the probability distribution of the gyration radius,
$R$, the two-point correlation functions and the average number of
intramolecular contacts as a function of $R$.

The scaling theory was then extended to treat the effects of interfaces and
geometrical restrictions. This direction of research is strongly motivated by
important applications to colloid stabilization and flocculation, liquid
chromatography, osmotic-pressure chromatography, ultrafiltration and others. A
number of problems are thereby closely related to the adsorption of polymers at
surfaces and interfaces. However, even in the absence of adsorption, geometric
constraints imposed on a polymer induce dramatic changes in its behavior which
have consequences pertaining to important applications. The basic model that
captures almost all the essential physics is a single polymer chain in a slit or
a tube with neutral (repulsive)
walls~\cite{Daoud,Brochard,Webman,Kremer,Cifra,Bleha,AMWPKB,Thompson,
Burkhardt,Wang,AMKB,Hagita,deJoannis,CifraBleha,Sotta,Bishop,Victor,Sheng,
CifraTeraoka,CifBle,CifTer,Hsu,Grassberger}. Lately,
the main focus has shifted to semidilute and concentrated
solutions~\cite{CifraTeraoka,TeraokaCifra,TeraokaWang}. Nevertheless, even the
basic model has intriguing properties some of which have not been fully
clarified yet.

The presence of a solid impenetrable wall gives rise to depletion
effects that were successfully treated within the scaling
framework~\cite{Eisenriegler,deGennes3,Eisen,deGennes4} and found experimental
verification. The most dramatic effects are observed when long polymer chains
are confined in slit-like or tube-like nanochannels where the chains become
effectively two- or one-dimensional. This situation is typical for biological
objects in living matter. Scaling theory formulated by de
Gennes~\cite{deGennes2,Daoud,Brochard} describes the basic polymer
characteristics, namely the average size of a chain and its confinement free
energy. It is well understood that the scaling expressions are valid only
asymptotically: $R\gg  D \gg a$, otherwise corrections to scaling come into
play. The scaling predictions have been successfully tested~\cite{Hsu,CifTer} by
means of Monte-Carlo simulation methods~\cite{Grassberger2}.

In this paper we present the results of detailed Molecular Dynamics (MD)
simulations using a coarse-grained bead-spring model with the aim of
establishing the range of validity of the scaling theory for uncharged flexible
polymers in good solvent strongly confined in a slit. One of the goals of the
paper is to give simple expressions that provide numerically accurate values for
all the essential characteristics of the chain with regard to the polymerization
index, $N$, and the slit width $D$.

We analyse the principle conformational parameters (end-to-end distance $r$
and the gyration radius $R$) for both lateral and normal directions. We examine
the probability distributions $W(r)$ and $W(R)$ for the lateral components of
$r$ and $R$, establish their universal character and provide simple analytical
expressions for them. Special attention is payed to the properties in {\em
normal} direction which have been largely ignored in earlier studies. Thus we
derive analytic expressions for the normal components of $\langle r^2\rangle$
and $\langle R^2\rangle$ which turn to be model-independent, in perfect
agreement with the simulation data. The  bond orientation profile across the
slit as well as the mean bond orientation against slit width $D$ have been
predicted analytically and verified here by computer experiment. The average
force produced by the polymer on the confining walls is calculated directly,
suggesting thus a scaling form for the free energy of confinement with accurate
numeric coefficients.

An essential part of our investigation examines dynamical properties of
polymers in confinement. We calculate the time autocorrelation functions for
the gyration radius and the end-to-end distance, extract the relaxation times,
and demonstrate distinct dynamic scaling behavior for lateral and normal
relaxation aa well as their mutual interplay.

Recently discovered possibilities of manipulating individual chains by AFM,
optical tweezers and surface force apparatus have generated strong interest in
the properties of end-grafted chains subject to external forces and confinement
effects~\cite{Clausen,Williams}. In view of these developments we present a
theoretical analysis and MD simulation data on the effects of grafting with
respect to the  static and dynamic characteristics of the polymer.

The paper is organized as follows: In Section \ref{model} we give a brief
description of the model, and in Section \ref{equil} we discuss the properties
of a confined chain in equilibrium. We give a derivation of the bond
orientation profile across the slit in the Appendix. Section \ref{dyn}
examines the relaxational and diffusive dynamics of the polymer. Finally we
discuss the implications of our results in Section \ref{discuss}.

\section{Model and Method}\label{model}

\begin{figure}[htb]
\includegraphics[scale=1.1]{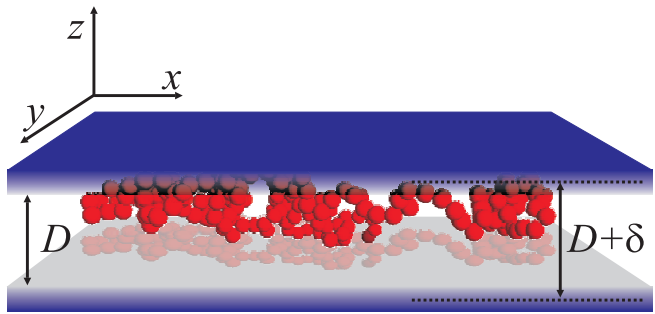}
\caption{A snapshot of a tethered chain with $N=200$ monomers in a slit of width
$D=4.0$. The soft repulsion of the wall potential is indicated by a gradient in
the color intensity. The {\em effective} width $D+\delta$ is indicated by
dashed lines. The grafting monomer is indicated by darker color.\label{snap}}
\end{figure}

The setup of our simulation is displayed in Figure~\ref{snap}.
The polymer chains are described by a simple coarse-grained bead-spring model,
originally proposed by Kremer and Grest~\cite{Grest}, which has been widely
and very successfully used for MD simulations of polymers in various
contexts~\cite{KB,Kotel}. Effective monomers along the chain are bound together
with a
 finitely-extensible non-linear elastic (FENE) potential,
\begin{equation}\label{FENE}
 U_{FENE}({\bf r})=-15\epsilon_w ({\cal R}_0/\sigma)^2 \ln\left (1- {\bf
r}^2/{\cal R}_0^2\right ),\;{\cal R}_0=1.5\sigma
\end{equation}
where $\sigma$ is the range parameter of a purely repulsive Lennard-Jones (LJ)
potential, that is truncated and shifted to zero in its minimum and acts
between any pairs of monomers.
\begin{equation}\label{LJ}
 U_{LJ}({\bf r})=4\epsilon
\left[(\sigma/{\bf r})^{12}-(\sigma/{\bf r})^6+1\right], {\bf r}\le
{\bf r}_c=2^{1/6}\sigma
\end{equation}
The parameter $\epsilon$, characterizing the strength of this potential, is
chosen unity, also the temperature $k_BT\equiv 1$, thus this potential gives
rise to an excluded volume interaction betwen all the non-bonded monomers of
the chain; for the bonded monomers, together with $U_{FENE}({\bf r})$, it leads
 to a typical neighbor - neighbor distance of ${\bf r}_{min}=0.96\sigma$.
Henceforth we choose $\sigma = 1$ as our unit of length.

Similarly, the confining walls are presented also by a purely repulsive wall
defined by Eq.~\ref{LJ} whereby the wall position is placed at the potential
minimum.

Molecular Dynamics (MD) simulations were performed using the standard
Velocity-Verlet algorithm \cite{Allen}, carrying out typically $1.5\times 10^9$
time steps with an integration time step $\delta t = 0.01 t_0$ where the MD time
unit (t.~u.) $t_0= (\sigma ^2m/48\epsilon_{LJ})^{1/2}=1/\sqrt{48}$, choosing the
monomer mass $m=1$. The temperature was held constant by means of a standard
Langevin thermostat with a friction constant $\zeta_0 = 0.5$

\section{Global Equilibrium Characteristics}
\label{equil}
\subsection{Free Energy and Force}

According to the blob picture~\cite{deGennes2}, a chain confined in a narrow
slit of width $D$ will form a two-dimensional self-avoiding walk consisting of
$n_b$ blobs of size $D$. Each blob contains $g=(D/a)^{1/\nu}$ monomers so that
the number of blobs is $n_b = N/g =N(D/a)^{-1/\nu}$ where $a$ is the distance
between neighboring monomers, and $\nu=0.58758(7)$~\cite{Hsu} is the Flory index
in $3$-dimensional space. The free energy excess of the confined chain (in units
of $k_BT$) is simply the number $n_b$ of blobs,
\be
   F_{\rm conf} =B n_b = B N (D/a)^{-1/\nu},     \label{Fconf}
\ee
where $B$ is a model-dependent dimensionless numerical coefficient. Note that
this free energy is extensive in $N$.

The confinement free energy and the associated force acting on the walls were
studied\cite{Hsu,Grassberger} by means of the pruned-enriched Rosenbluth method
(PERM algorithm) that allows one to estimate the partition
function~\cite{Grassberger2}. For the off-lattice model of Ref.~\cite{AMKB}, the
pressure tensor was calculated using the virial theorem and studied in
conjunction with the normal density profile. For relatively narrow slits the
simulation data for the force were consistent with the scaling prediction
$f\approx N D^{-1-1/\nu}$.

For self-avoiding walks on a simple cubic lattice the confinement free energy
was calculated by MC methods in \cite{CifBle,Hsu}. These results  were presented
by a general power-law fit as $F_{\rm conf} = 0.843 N^{0.94} (D/a)^{-1.57}$.
According to this fit, the free energy is not exactly extensive and the
$D$-exponent deviates from the scaling prediction which makes the fitting
formula purely phenomenological. Hsu and Grassberger~\cite{Hsu} have obtained
accurate data for very long chains ($N$ up to $8000$ and slit width up to
$D=120$) and verified the scaling prediction. They also noted that there must be
a non-universal correction to the value of the slit width $D$ to $D+\delta$ with
$\delta =0.33$, since the original scaling formula assumes the asymptotic limit
$D\gg a$. Their best fit for the confinement free energy was $F_{\rm conf} =
2.10 N (D/a)^{-1/\nu}$. The importance of proper corrections to the slit width
in order to achieve good scaling was emphasized by Milchev and
Binder~\cite{AMKB} and by Teraoka~\cite{CifTer}.

The MD method does not allow to obtain the free energy, but the force acting on
the walls can be calculated directly.The force acting on the walls is directly
derived from the confinement free energy
\be
   fa = - d F_{\rm conf}/dD = {(B/\nu)} N (D/a)^{-1-1/\nu}     \label{f}
\ee
\begin{figure}[htb]
\includegraphics[scale=0.9]{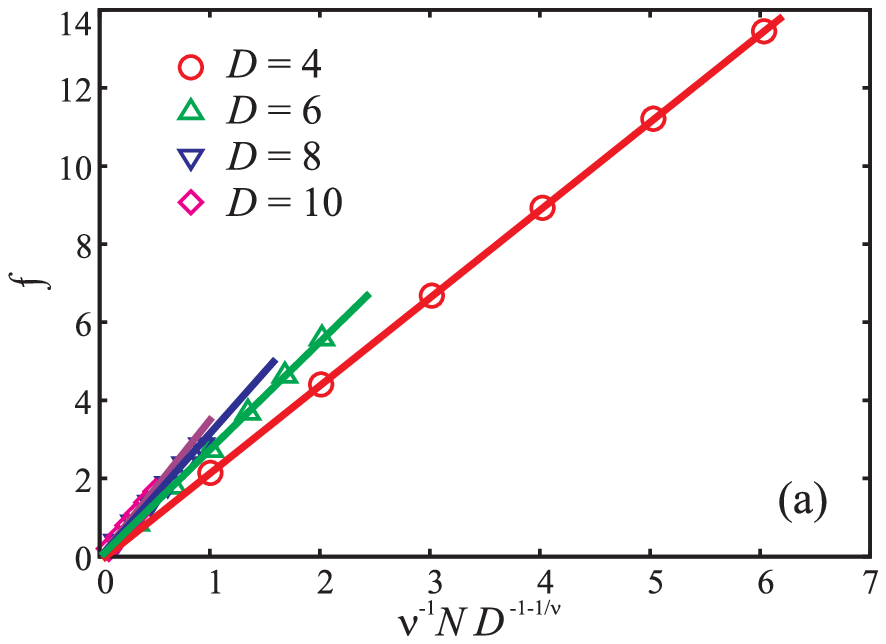}
\includegraphics[scale=0.9]{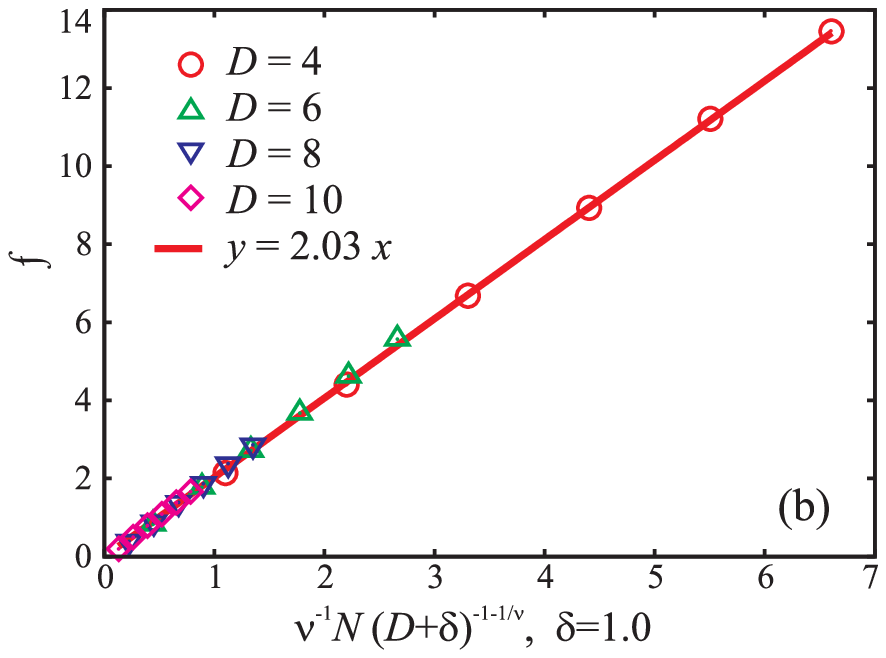}
\caption{Variation of the force $f$, exerted by a confined polymer chain on the
slit walls, with scaling variable $\nu^{-1}ND^{-1-1/{\nu}}$: (a) without,
and (b) with taking the width correction $\delta$ into account.\label{force}}
\end{figure}

Figure ~\ref{force} presents the force against the scaling combination  $N
(D/a)^{-1-1/\nu}$ with and without a shift $\delta$. It is clear from Figure
~\ref{force} that the data points do not fall on a universal curve, if
$\delta=0$. The best universal fit for the force was obtained for $\delta=a$:
\be
   fa = 2.03 \nu^{-1}ND^{-1-1/{\nu}} .
\label{fitf}
\ee
 The corresponding free energy is given by
 \be
   F_{\rm conf} = 2.03 N (D/a+1)^{-1/\nu}     \label{fitFconf}
\ee
This correction to $D$ turned out to be relevant not only for the free energy
but for describing the variation of all other quantities as functions of $D$
too. Throughout the paper we use the same slit width correction $\delta=a$. It
is remarkable that the scaling asymptotic relation for the confinement free
energy is achieved very early, when the number of blobs is still close to unity:
the minimal number of blobs in our simulations was $n_b=N(D/a+1)^{-1/\nu}=0.84$
for $N=50$ and $D=10$.

Comparing our results for the free energy with those obtained for a lattice
model we notice that a very good semi-quantitate estimate of the slit
confinement free energy is $2k_BT$ per blob. It is also quite remarkable that
the same estimate~\cite{Grassberger} works very well for the free energy of a
$2D$ chain confined in a strip ($ F_{\rm conf} = 1.944 n_b$ where $n_b
=ND^{-1/\nu_2}$).

\subsection{Density profiles}

\begin{figure}[htb]
\includegraphics[scale=0.88]{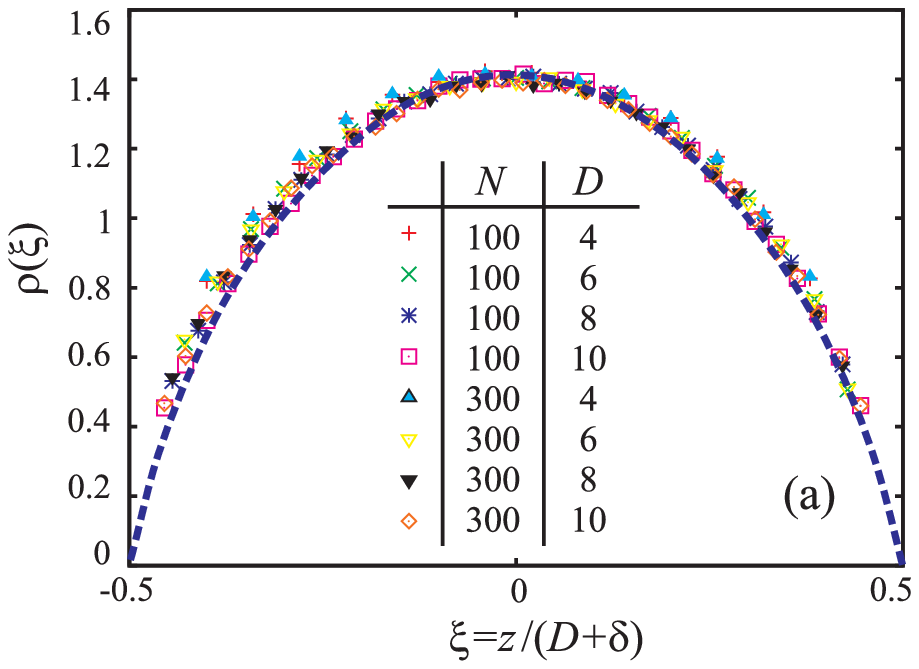}
\includegraphics[scale=0.88]{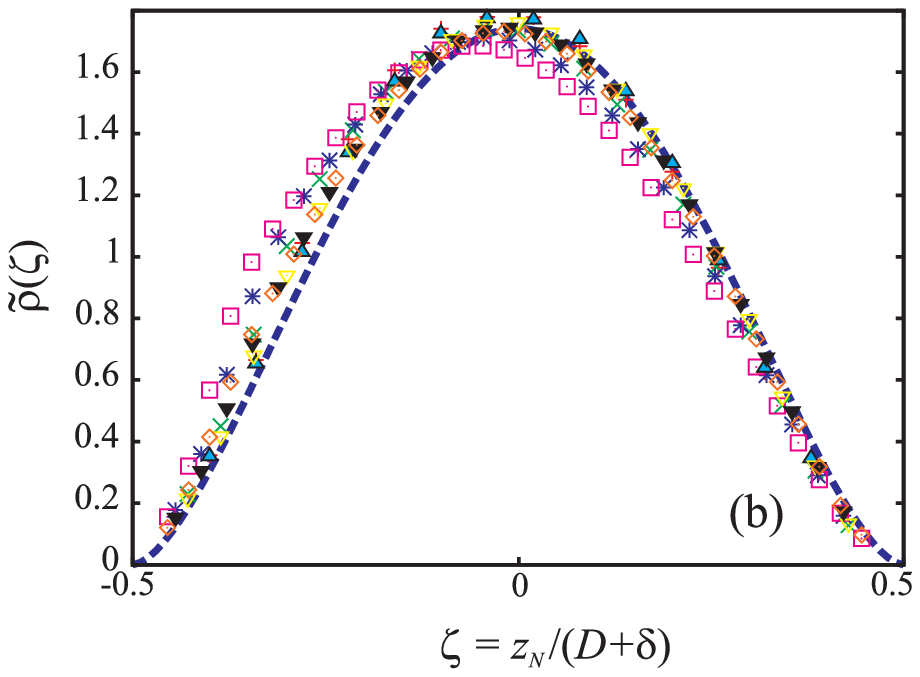}
\caption{Density distribution of end-monomers (a), and all monomers (b), across
the slit width for tethered chains of different length $N$ and slits of
different width $D$ (both given as parameters). Dashed lines denote the
theoretical results, see Eqs.~\ref{End distr}, \ref{Full distr}.\label{density}}
\end{figure}

The distribution of the end monomer position across the slit was studied
extensively by Hsu and Grassberger~\cite{Hsu}. The scaling variable is taken as
the ratio $\xi=z/D$ and a simple scaling formula that takes into account the
mirror symmetry with respect to the median plane was proposed. If the coordinate
$z$ is counted from the median plane, the expression has the following form
\be
   \rho_{\rm end}(\xi) =
\frac{\Gamma(2+1/\nu)}{\Gamma^2(1+1/2\nu)}(1/4-\xi^2)^{1/2\nu}
\label{End distr}
\ee
The distribution Eq.~\ref{End distr} is properly normalized to unity. It is well
known that for an ideal chain in a slit the total monomer density profile is
proportional to the square of the end segment distribution, as long as the
strong confinement condition is satisfied. A hypothesis that the same relation
holds also for self-avoiding chains was checked by Hsu and
Grassberger~\cite{Hsu} and good agreement was observed. With this ansatz, the
monomer density profile can be written as
\be
   \rho(\xi) =\frac{\Gamma(2+2/\nu)}{\Gamma^2(1+1/\nu)}(1/4-\xi^2)^{1/\nu}
\label{Full distr}
\ee
Density distributions of end-monomers and of all monomers across
the slit width (for tethered chains of different length $N$ and slits of
different width $D$) are presented in Figure~\ref{density} together with the
theoretical curves. The data points for various values of $N$ and $D$
collapse onto the universal theoretical curves containing no fitting parameters.
Some deviations violating the mirror symmetry can be seen on the full
density distribution picture. These are due to the effects of grafting which are
especially noticeable for the shortest chain $N=100$ in a wide slit $D=10a$.
The fact that the analytical formulas based on the scaling ansatz work very
well both for the end monomer distribution and the full density profile for
lattice as well as for off-lattice models is quite remarkable.

The scaling ansatz has been also verified by Hsu and
Grassberger~\cite{Grassberger} for two-dimensional chains (with $\nu_2=3/4$)
confined in a strip, and for confined ideal chains (with $\nu_G=1/2$) in a slit.
We would confidently speculate that the formulae will work for real flexible
polymers in good and $\theta$-solvents including exotic situations when strong
adsorption enforces essentially 2D conformations while additional repulsive
barriers confine the chain inside a strip.

\subsection{Average chain size characteristics.}

\begin{figure}[htb]
\includegraphics[width=8cm, height=6cm]{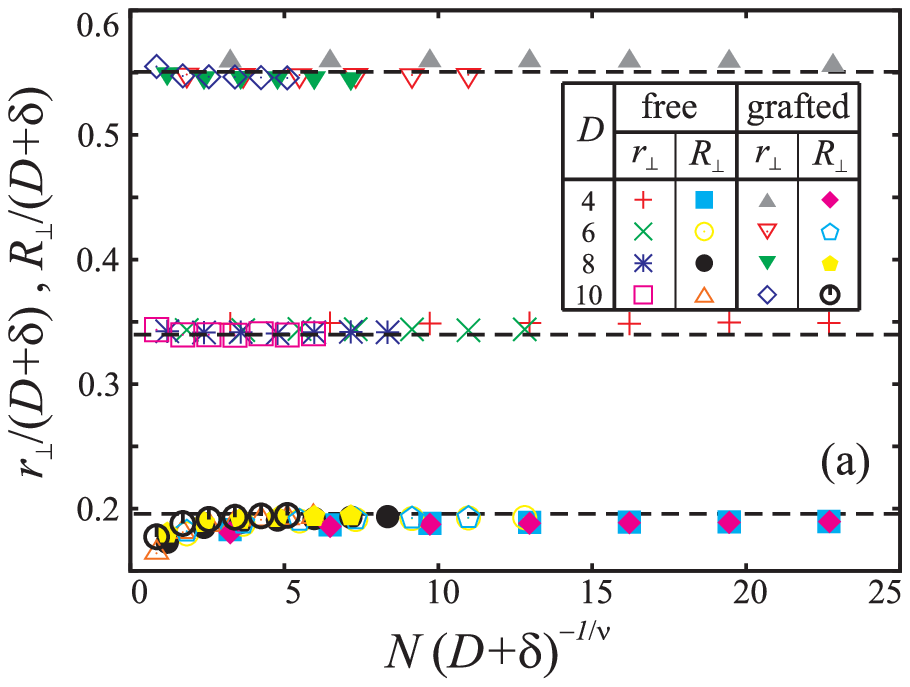}
\includegraphics[width=8cm, height=6cm]{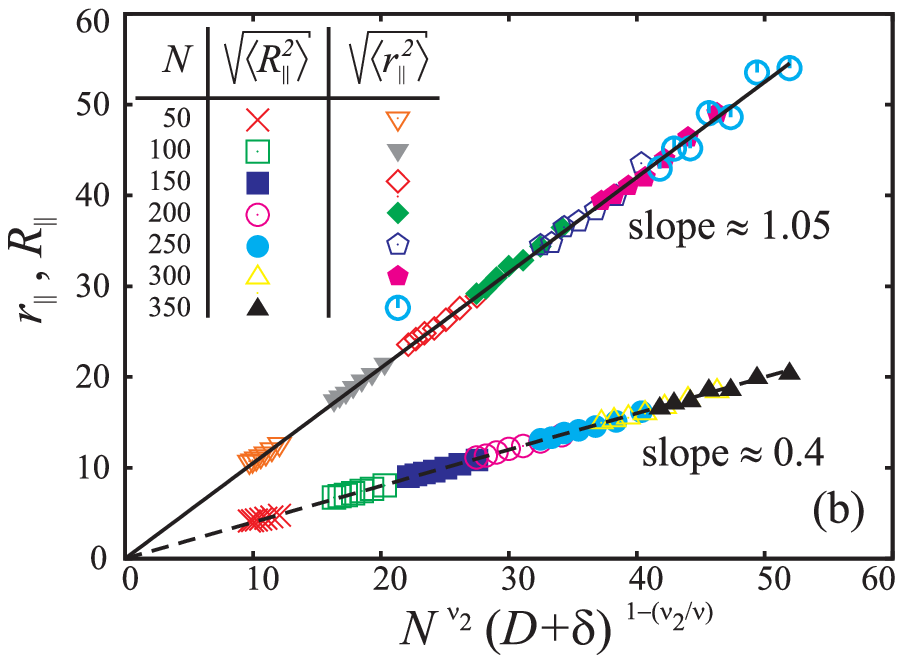}
\caption{ (a) Normal components of the end-to-end vector,
$r_{\perp}$, and the gyration radius, $R_{\perp}$, against the number of blobs
$N(D+\delta)^{-1/\nu}$. (b) Variation of the parallel components of the
end-to-end vector, $r_{\parallel}$, and gyration radius, $R_{\parallel}$, with
scaling variable $N^{\nu_2}(D+\delta)^{1-\nu_2/\nu}$ for chains of length $N$
and different slit width $4\le D\le 10$.\label{perp_size}}.
\end{figure}

The characteristics for a polymer chain  average size can be reduced to moments
of the joint monomer-monomer distribution $\hat{\rho}(z_i,z_j)$. In
particular, the mean square of the gyration radius $\langle R_\perp^2\rangle$
is calculated as:
\be
  \langle R_\perp^2\rangle =
(1/2N)\sum_{i,j=1}^N\int\int(z_i-z_j)^2\hat{\rho}(z_i,z_j)dz_idz_j
  \label{Rperp}
\ee
whereby the integration in Eq.~\ref{Rperp} is carried out across the slit.
It follows naturally from the scaling blob picture of a chain strongly confined
in a slit that the joint distribution splits into a product of single-monomer
distributions
$\hat{\rho}(z_i,z_j)={\rho}(z_j){\rho}(z_i)$ as
long as the monomers are separated along the chain by at least one blob,
$|j-i|\> D^{1/\nu}$. We assume that the dominant contribution comes from
the internal monomer pairs (that is, pairs which do not belong to the two
terminal blobs). Hence the single-monomer distribution ${\rho}(z_j)$ is
independent of the index $j$ and coincides with the normalized full density
profile $\rho(z)$. It follows that
\begin{eqnarray}
\langle R_\perp^2\rangle = \frac{1}{2}\int\int (z-z')^2{\rho}(z){\rho}(z') dz
dz' \nonumber \\
 = \left [ \frac{1}{4} +
\frac{\Gamma(2+\frac{2}{\nu})}{\Gamma(1+\frac{1}{\nu})} \left(
\frac{\Gamma(3+\frac{1}{\nu})}{\Gamma(4+\frac{2}{\nu})} -
\frac{\Gamma(2+\frac{1}{\nu})}{\Gamma(3+\frac{2}{\nu})}\right)\right] D^2
\label{Rperp answer}
\end{eqnarray}
i.~e., $\langle R_\perp^2\rangle = 0.04D^2$, where the scaling form of the
full density was used. The derivation suggests that effects of grafting should
not be discernable in $\langle R_\perp\rangle$. The average square of the normal
end-to-end
distance is defined as
\be
\langle r_\perp^2\rangle = \int \int (z_1-z_N)^2 \hat\rho(z_1,z_N)dz_1 dz_N
\label{r perp}
\ee

We employ the same factorization assumption to separate the probability
densities for the two chain ends. In the case of a non-grafted chain both ends
are described by $\rho_{\rm end}(z)$ distributions leading to

\begin{eqnarray}
\langle r_\perp^2 \rangle
&=& \int\int(z_1-z_N)^2\rho_{end}(z_1)\rho_{end}(z_N)dz_1dz_N  \nonumber \\
&=& \left [ \frac{1}{2} +
\frac{2\Gamma(2+\frac{1}{\nu})}{\Gamma(1+\frac{1}{2\nu})} \left(
\frac{\Gamma(3+\frac{1}{2\nu})}{\Gamma(4+\frac{1}{\nu})} -
\frac{\Gamma(2+\frac{1}{2\nu})}{\Gamma(3+\frac{1}{\nu})}\right)\right] D^2
\nonumber \\
&=&0.1D^2.
\label{r perp ans}
\end{eqnarray}

If one end is fixed (tethered) at some point $z_{gr}$, then the distribution
for this end is given by ${\rho}(z_1)=\delta(z_1-z_{gr})$ reducing the
formula to $\langle r^2_\perp(z_{gr}) \rangle = \langle r^2_\perp(0)
\rangle + z_{gr}^2$ where
\be
    \langle r_\perp(0)^2 \rangle = \left (
\frac{\Gamma(3+\frac{1}{2\nu})\Gamma(2+\frac{1}{\nu})}
{\Gamma(4+\frac{1}{\nu})\Gamma(1+\frac{1}{\nu})} - \frac{1}{4}
\right) D^2  \label{r perp graft}
\ee

The result is quite general and covers a whole class of confined systems
characterized by different values of the Flory index $\nu$. In our
model the chain end was fixed at $z_{gr}=D/2-1$ and the  simulation data are
to be compared to the theoretical expression
\be
 \langle r_\perp^2 \rangle_{gr} = 0.3D^2 - D + 1
\label{r perp graft ans}
\ee
The formula suggests that the combination
\be
  D^{-1}\sqrt{\langle r_\perp^2 \rangle_{gr}+D-1}    \label{r perp graft
corrected}
\ee
remains constant independent of $D$ and $N$. Strictly speaking, the above
theoretical predictions are valid only asymptotically for $n_b\gg 1$. The
simulation data for the three characteristics calculated above are presented in
Figure \ref{perp_size}a as functions of the blob number. The end-to-end distance
for the grafted chains is corrected according to Eq.~\ref{r perp graft
corrected}. In full agreement with the asymptotic theory, the normalized values
are independent of $D$ and $N$, and the actual values coincide with the
theoretical predictions. Small deviations are visible only for the gyration
radius curve with number of blobs close to $1$. It is quite remarkable that the
limiting asymptotic values are achieved even for such  moderate compression.

It is well known that the size of a free chain in the bulk is proportional to
$N^\nu$. For the Kremer-Grest model~\cite{Grest} that we use in our MD
simulations, the model-dependent numerical prefactors defining the rms average
of the gyration radius and the end-to-end distance have been estimated as
\be
     {R_F} = {0.27 a N^{\nu}}\;\mbox{and}\; {r_F} = {0.67 a N^{\nu}}.
 \label{Rfree}
\ee
According to the blob picture, a chain in a slit is a two-dimensional
self-avoiding walk consisting of $n_b$ blobs of size $D$. Thus scaling predicts
the average lateral size to be
\be
 R_\|\sim n_b^{\nu_2}D=aN^{\nu_2}(D/a)^{1-\nu_2/\nu}.
\label{rh}
\ee
If $D$ is of the same order as $R_F$, Eq.~(\ref{rh}), one finds $R_\|\sim a
N^{\nu}$, providing the expected smooth crossover.

Figure~\ref{perp_size}b presents the average lateral  $r_\|$ component of the
end-to-end distance and the gyration radius versus the scaling arguments as
suggested by the theory. For the lateral size, no effect of grafting was
observed.

The best fit for the average lateral end-to-end distance is
\be
        r_\|=1.05 aN^{\nu_2}(D/a+1)^{1-\nu_2/\nu}          \label{r_0}
\ee
and for the average lateral component of the gyration radius

\be
        R_\| =0.40 aN^{\nu_2}(D/a+1)^{1-\nu_2/\nu}          \label{R_g}
\ee
Although the amplitudes are non-universal, it would have been instructive to
compare them to those of other models in order to see the variation range.
Unfortunately, only one amplitude for the end-to-end distance of chains on a
cubic lattice $ r_\|=0.835 aN^{\nu_2}(D/a+1)^{1-\nu_2/\nu} $ could be found
in the literature~\cite{Grassberger}.

\subsection{Probability distributions of end-to-end distance and radius of
gyration}

The scaling form for the end-to-end vector distribution of a self-avoiding
polymer chain was suggested by Fisher~\cite{Fisher} in 1966 and then refined by
Cloizaux~\cite{Cloizeaux} and others. In the MD simulations we obtain the
histogram for the scalar end-to-end distance. The distribution for the reduced
scalar distance ${\eta=(r/r_0)}$ in $d$ dimensions can be represented as
\be
   W_d(\eta) = A r_0^{-1} \eta^{\left(d-1\right)} \eta^{\theta_d}
exp(-B\eta^{\delta_d})    \label{W end dist}
\ee
where $r_0$ is the average end-to-end distance, and A and B are numerical
coefficients. The exponent $\delta_d=1/(1-\nu_d)$ describes strong stretching of
the chain and is related to the Flory exponent $\nu_d$ for a chain in the
$d$-dimensional space. The exponent $\theta_d=(\gamma_d-1)/\nu_d$  is related to
another critical exponent $\gamma_d$ that appears in the partition function
$Q_N= \mu^{-N}N^{\gamma_d -1}$ where $\mu$ is called an {\em effective}
coordination number. In $d=2$ one has $\gamma_2=43/32$~\cite{deGennes}, in $d=3$
the best estimate is $\gamma_3 = 1.1575(6)$~\cite{Caracciolo}. The extra factor
$r_0^{-1} \eta^{\left(d-1\right)}$ that appears in Eq.~\ref{W end dist} in
comparison to the cited references is due to the volume element included in the
definition of the distribution of the scalar distance. According to this
definition, the two normalization conditions have the following form: $\int
W_d(r)dr=1$ and $\int rW_d(r)dr=r_0$, and they fix uniquely the vales of $A$ and
$B$:

\be
   B
=\left[\Gamma\left(\frac{d+1+\theta_d}{\delta_d}\right)/\Gamma\left[\frac{
d+\theta_d)}{\delta_d}\right)\right]^{\delta_d}     \label{B}
\ee
\begin{eqnarray}
   A
&=&\frac{\delta_d}{\Gamma\left(\frac{d+\theta_d}{\delta_d}\right)}\left[
\Gamma\left(\frac{d+1+\theta_d}{\delta_d}\right)/\Gamma\left[\frac{d+\theta_d)}{
\delta_d}\right)\right]^{(d+\theta_d)} \nonumber \\
&=&1.374     \label{A}
\end{eqnarray}

The final form of $W_d(\eta)$ for $d=2$ is

\be
   W_2(\eta) = 1.374\eta^{1.458} exp\left(-0.324\eta^{4}\right)
\label{W2}
\ee

For $d=3$ the distribution is given by

\be
   W_3(\eta) = 3.032\eta^{2.268} exp\left(-1.084\eta^{2.425}\right)
\label{W3}
\ee

It is clear that the distribution $W(\eta)$ for a finite chain in a slit has to
be between $W_2$ and $W_3$. For narrow slits and long chains, $r_0\gg D$, the
distribution $W(r)$ is expected to be close to $W_2$. For wide slits or short
chains with  $r_0\approx D$, $W(\eta)$ should be closer to the
three-dimensional distribution $W_3$.

\begin{figure}[htb]
\includegraphics[scale=0.9]{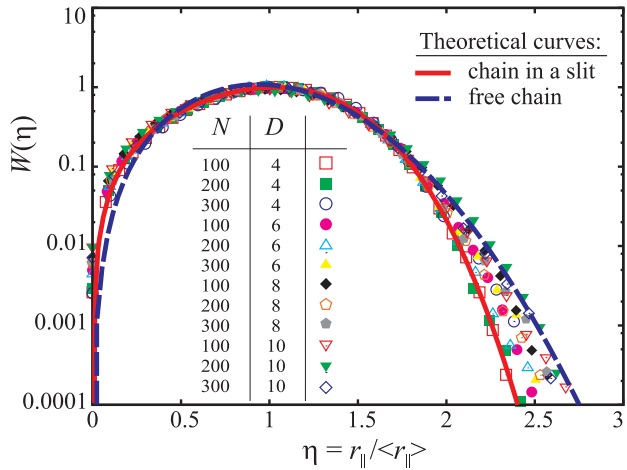}
\caption{Probability distribution functions of the end-to-end distance of a
polymer chain parallel to the slit wall for varying chain length $N$ and slit
width $D$. Lines indicate the analytic results for a confined chain (full
line), Eq.~\ref{W2}, and for a chain without geometric
constraints, Eq.~\ref{W3}, (dashed line). The curves are normalized so that
the average value $\langle \eta\rangle = 1$ and $W(\langle
\eta\rangle)=1$.\label{PDF_end}}
\end{figure}

In Figure ~\ref{PDF_end} we plot the  distribution $W(\eta)$ of the lateral
component of the end-to-end distance (in semi-log scale) against $\eta=r/r_0$
for various chain lengths $N$ and slit widths $D$. The analytical equations
$W_2$ and $W_3$ for chains in two-dimensions and three-dimensions are shown by
solid and dotted lines, respectively. In the region around the maximum both
distributions are close to each other and all data points collapse on a
universal curve. For large extensions $r>1.8 r_0$, the strongly confined chains
follow the two-dimensional curve $W_2$ while for chains in relatively wide slits
the data points lie closer to the three-dimensional curve $W_3$. In this region
$W(\eta)$ is not universal and depends on the ratio $r_0/D$. The distribution of
the gyration radius for a chain in a narrow tube  was postulated by
Victor~\cite{Victor} in scaling form, and verified numerically by Sotta et
al.~\cite{Sotta} and Bishop et all.~\cite{Bishop}. They also calculated the
probability distribution for the end-to-end distance and fitted their results
by a similar expression.The distribution of the gyration radius and of the
end-to-end distance for a chain in a slit has not been analyzed yet.
\begin{figure}[htb]
\includegraphics[scale=0.9]{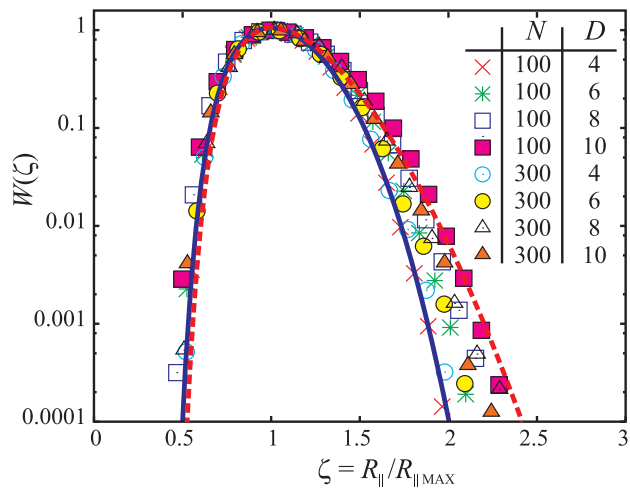}
\caption{Probability distribution functions of the gyration radius
$R_{\parallel}$, parallel to the slit wall. Here curves are normalized so that
the maximum value $\zeta_{MAX} = 1$ and $W(\zeta_{MAX})=1$.\label{PDF_Rg}}
\end{figure}

In Figure ~\ref{PDF_Rg} we plot the  distribution $R_{\parallel}$ of the lateral
component of the gyration radius (in semi-log scale) vs $\zeta=R_|/R_0$ for
various chain lengths $N$ and slit widths $D$. The analytical equations $W_2$
and $W_3$
\be
   W_2(\zeta) = 0.65\left( \zeta^{-4} +\zeta^{4}-2\right)
\label{W2_Rg}
\ee
and
\be
   W_3(\zeta) = 1.34\left(\zeta^{-4} +\zeta^{4}-2\right)  \label{W3_Rg}
\ee
for two-dimensional and three-dimensional chains are shown by solid and dotted
lines, respectively.  In the region around the maximum both distributions are
close to each other and all data points collapse on a universal curve. For large
extensions $R>1.8 R_0$, the strongly confined chains follow the two-dimensional
curve $W_2(\zeta)$ while for chains in relatively wide slits the data points lie
closer to the three-dimensional curve $W_3(\zeta)$. In this region $W(\zeta)$ is
not universal and depends on the ratio $R_0/D$.

\subsection{Confinement effect on segment orientation}

A commonly used measure characterizing the orientation of bonds is the average
\begin{figure}[htb]
\includegraphics[scale=0.88]{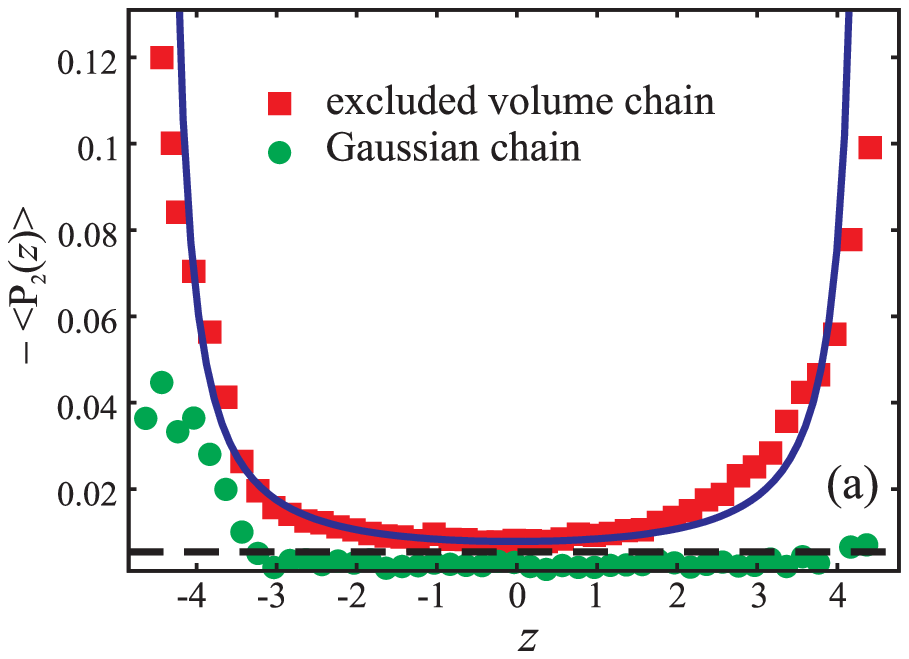}
\includegraphics[scale=0.88]{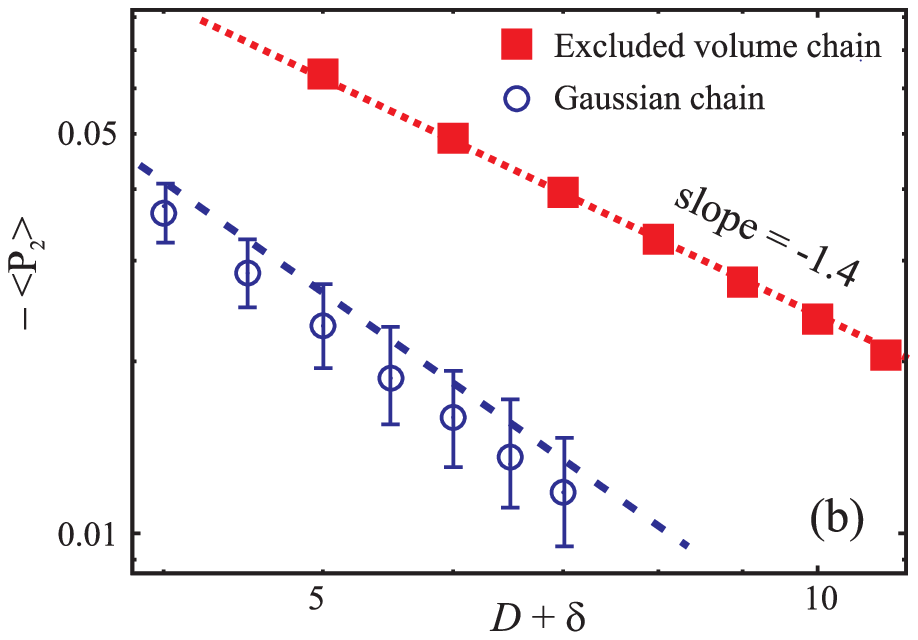}
\caption{(a) Profile of mean bond orientation $\langle P_2(z)\rangle = 2^{-1}
\left [3\langle \cos^2\theta\rangle - 1\right ]$ across the slit where the
orientation of the bonds is measured with respect to the $Z$-axis, for a real
and Gaussian tethered chain. A horizontal dashed line denotes the theoretical
prediction, $\langle P_2(z)\rangle_{\mbox{id}} =
-\frac{\pi^2}{15}(a/D)^{2}$, while a full line marks the scaling
conjecture $\langle P_2(z)\rangle \propto (\frac{D^2}{4} - z^2)^{2(1-1/\nu)}$.
(b) Variation of $\langle P_2\rangle$, averaged over $z$, against the effective
slit width $D+\delta$. Circles denote simulation results for a Gaussian chain,
squares - for a chain with excluded volume interactions. A dotted straight line
indicates the theoretically predicted slope of $-1.4$, and a dashed line shows
the theoretical result for $\langle P_2\rangle_{\mbox{id}}$,
Eq.~\ref{pi}.\label{Legendre}}
\end{figure}
value of the second Legendre polynomial of the azimuthal angle,  $\langle
P_2\rangle=2^{-1}(3\langle \cos^2\theta\rangle -1)$. Experimentally this
parameter appears in NMR and optical birefringence measurements. In a free
non-confined coil the bond orientation is completely isotropic, and $\langle
P_2\rangle=0$. For a chain in a slit, it is natural to expect a preferential
bond orientation along the lateral plane which would lead to a non-zero negative
value of $\langle P_2\rangle$. In Figure~\ref{Legendre}a the profile of the
average orientation across the slit obtained in the MD simulation is displayed
together with the similar profile calculated for the same model but with the
excluded volume interactions between non-neighboring monomers switched off. The
values of $\langle P_2\rangle$ averaged over all monomers irrespective of their
position are plotted against the effective slit width in Figure~\ref{Legendre}b
in a log-log scale, the best fit for the slopes being $-2.0$ for the ideal
chain, and $-1.44$ for the 'real' chain (with excluded volume). It is clear that
the excluded volume interactions affect the magnitude of the orientation very
strongly.

A naive estimate of the orientation effect as a function of the slit width can
be obtained as follows: assume that the wall induces some lateral orientation
only locally, within a distance of the order of a monomer size $a$. The local
orientation is by itself independent of $D$. The fraction of monomers within
distance $a$ from the wall is given by $\rho(D/2+a)\sim D^{-1-1/\nu}$ which
would give $D^{-2.70}$ for a chain with excluded volume and $D^{-3}$ for an
ideal chain. Both estimates turn out to be well below the magnitude of the
observed effect. Next, we introduce the effect of {\em orientation correlations}
along the backbone of the chain. It is known that the simplest bond-bond
orientation correlation function $P_1$ for an ideal chain decreases
exponentially with the distance $s$ along the chain, $\exp(-\kappa s)$ where
$\kappa \propto 1/a$ for flexible chains. In a real chain the correlations decay
according to a power law $s^{-\omega}$ where $\omega=2(1-\nu)$~\cite{Obukhov}.
We will assume that the correlations that transmit the orienting effect of the
wall, as described by the $\langle P_2\rangle$ parameter, propagate along the
chain according to the same laws. In the spirit of the scaling theory, the
contour distance $s$ can be related to the normal distance from the wall, $z$,
as $s=z^{1/\nu}$. Thus the relationship $\langle P_2(z))\rangle \propto
z^{-\omega/\nu}$ can be represented in a normalized form as $\langle
P_2(z)\rangle \propto (\frac{D^2}{4} - z^2)^{2-2/\nu}$. Figure~\ref{Legendre}a
demonstrates that the simulation data appear in very good agreement with this
theoretical prediction.

By integrating over the slit width one obtains the average orientation
parameter for the real chain as
\be
   \langle P_2\rangle = \langle P_2\rangle_{wall}\int_{-D/2}^0	z^{-\omega/\nu}
\rho(z) dz = const. D^{-\omega/\nu}\label{Leg correl real}
\ee

A similar calculation for the ideal chain gives
\be
  \langle P_2\rangle = \langle P_2\rangle_{wall}\int_{-D/2}^0 e^{-\kappa z}\rho
_{id}(z) dz = const. D^{-3}.	  \label{Leg corr id}
\ee
It is clear that the scaling law obtained for the real chain, $\langle
P_2\rangle \sim D^{-2(1/\nu-1)}= D^{-1.40}$ agrees very well with the simulation
data while the theoretical result for the ideal chain $\langle P_2\rangle\sim
D^{-3}$ still underestimates the observed effect. This suggests that apart from
the local orienting effect of the wall there must be another mechanism inducing
preferential orientation in the lateral plane. A theory describing this
mechanism is given in the Appendix. The main result for the ideal chain is

\be
    \langle P_2\rangle_{\rm id} = -\frac{\pi^2}{15 D^2}     \label{Leg id}
\ee
which agrees well with the simulation data without any fitting parameters.

\section{Dynamic scaling}
\label{dyn}

There are only few simulation studies on the dynamics of a polymer chain
confined
in a slit. For such a chain only the center-of-mass diffusion has been
studied by MC for an off-lattice model~\cite{AM}. The characteristic diffusion
time was found to scale as $\tau\sim N^{5/2}$ which conforms to the dynamic
scaling prediction $\tau \propto R^2/N$ using the static scaling prediction
$R\sim N^{\nu_2}$. Here we present a study of relaxation dynamics for the
gyration radius, derived from the analysis of the appropriate autocorrelation
functions, and compare it to the center-of-mass diffusion characteristic time.
The time-dependent autocorrelation functions were calculated as follows:
\be
   \langle C_R(t) \rangle = \frac{\langle R(0)^2 R(t)^2\rangle - \langle
R(0)^2\rangle ^2}{\langle R(0)^4\rangle - \langle R(0)^2\rangle ^2}
 \label{corr func}
\ee
separately for the lateral and perpendicular components of the gyration radius.
\begin{figure}
\includegraphics[scale=0.89]{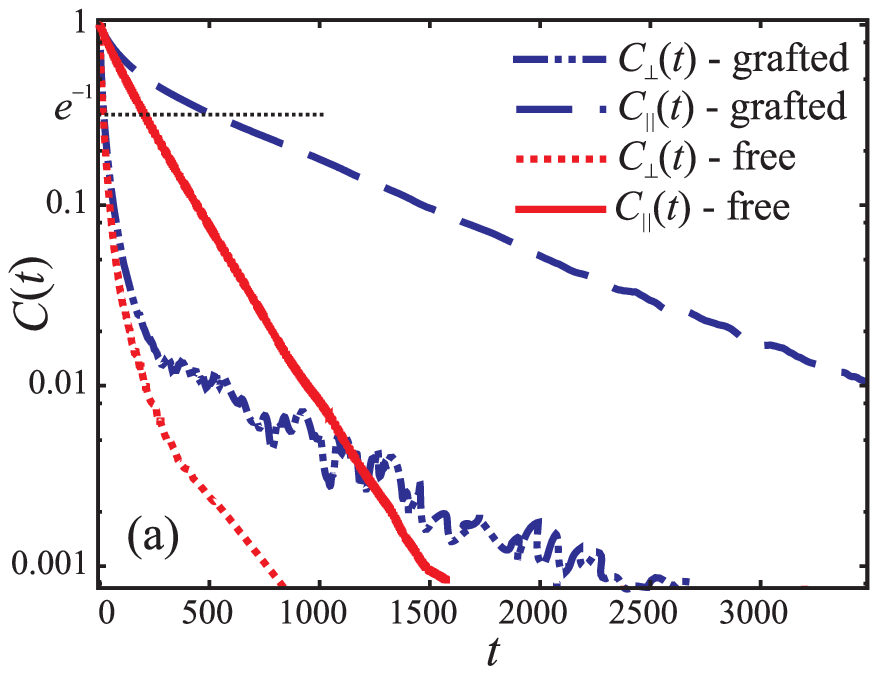}
\includegraphics[scale=0.89]{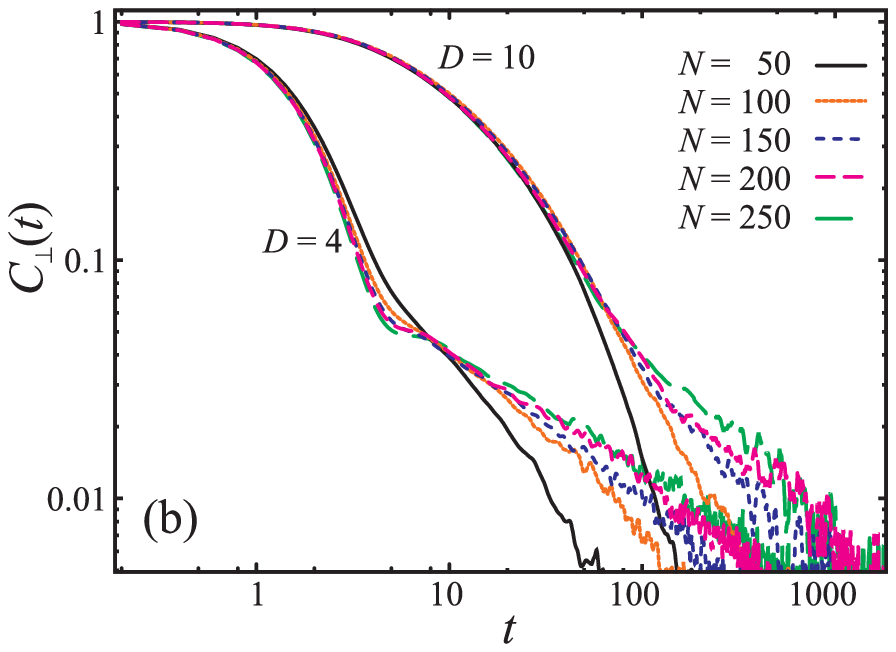}
\caption{(a) Comparison of the time autocorrelation functions (ACF) of the
normal and parallel gyration radius components for grafted and free chains with
$N=100$ in a slit with $D=10$. The intersection of $C_{\perp}(t)$ with the
horizontal dotted line at $e^{-1}$ has been used to determine the characteristic
time of the fast chain relaxation perpendicular to the slit wall, shown in
Fig.~8b. (b) Log-log plot of the ACF of the gyration radius normal
component for two slit widths $D=4$ and $D=10$ and different chain
lengths.\label{ACF}}
\end{figure}

Typical curves are presented in Figure~\ref{ACF}. Part (a) demonstrates the
effect of grafting one chain end. Evidently, this slows down considerably the
lateral relaxation. The grafting effect on the normal relaxation is more
complicated, and this is directly related to the shape of the ACF. The major
initial portion of normal ACF is characterized by a rapid decay with a
relaxation time {\em unaffected} by grafting. However, there is  clearly a
visible tail of rather small amplitude that is described by a much slower
relaxation. A comparison with the lateral ACFs suggests that we encounter a
classic example of weakly coupled degrees of freedom with a large difference in
their time scales. The observed slow tails of the transverse ACFs repeat exactly
the corresponding tails for the longitudinal relaxation. The semilog scale of
the figure allows one to see the dominant relaxation time of the longitudinal
ACFs although the ACFs are not perfectly straight and indicate some contribution
from the faster modes in their initial decay. The fundamental longitudinal
relaxation times can be estimated from the slopes of the linear part of the ACF
rather reliably. The situation with the trasverse relaxation, however, is
complicated by the fact that the initial relaxation contains also small-scale
contributions from different modes whereas the long-term relaxation is dominated
by coupling to lateral modes.

Part (b) of the Figure displays the normal ACFs for various values of $N$ and
two values of $D$ in a log-log scale. It is clear that the major initial portion
of the normal relaxation is $N$-independent although it depends strongly on the
slit width. On the other hand, the long tails demonstrate a dependence on both
$N$ and $D$ due to coupling, as one would expect for the longitudinal
relaxation.

The scaling theory for the longitudinal relaxation time is rather
straightforward and the main ideas were discussed before by
Descas et al.~\cite{Sommer}. The relaxation time scales as
\be
   \tau_{\|}\sim R^2 \zeta_{fr}  \label{tau r2}
\ee
where $\zeta_{fr}$ is the total chain friction coefficient. For a free-draining
chain, $\zeta_{fr}=N\zeta_0$ with $\zeta_0$ being the friction per one monomer.
It follows immediately that the relaxation time is proportional to a scaling
parameter
\be
\psi=N^{1+2\nu_2}(D+\delta)^{2(1-\nu_2/\nu)} \label{tau scaling}
\ee

\begin{figure}
\includegraphics[width=8cm, height=6cm]{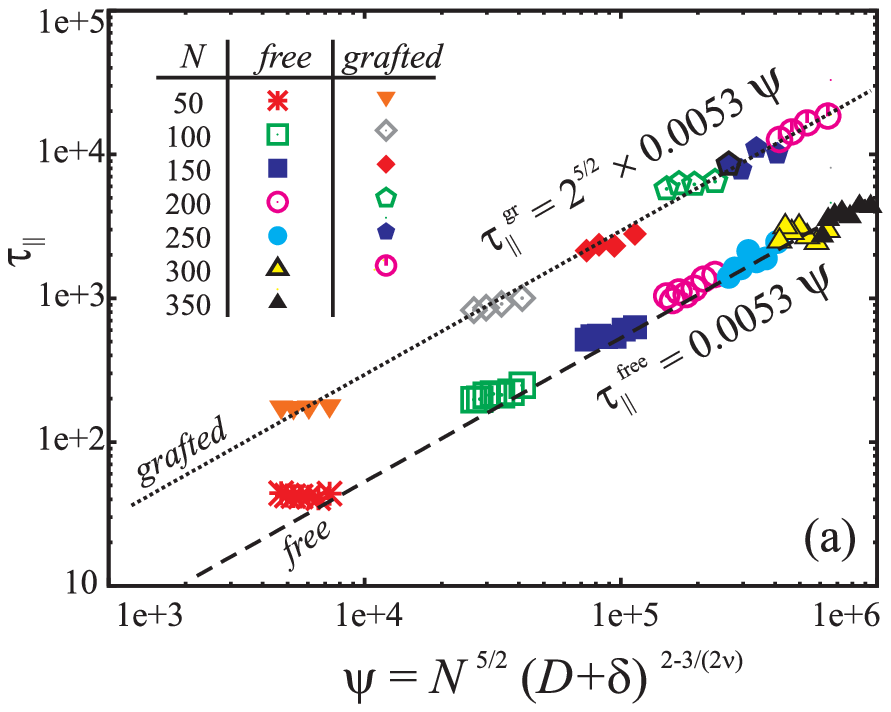}
\includegraphics[width=8cm, height=6cm]{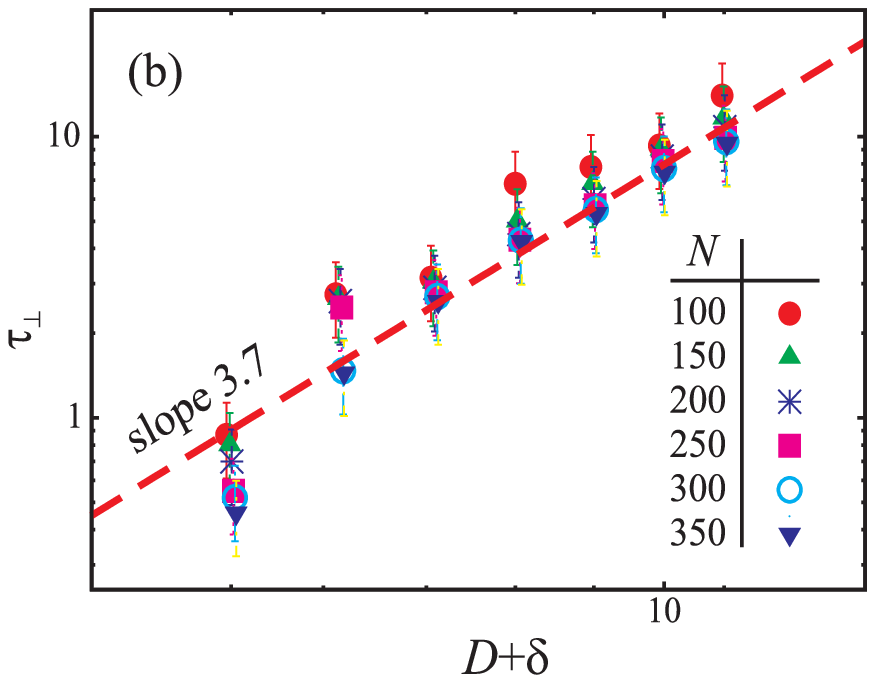}
\caption{(a) Mean relaxation times parallel to the slit wall for free and
tethered chains with different length $N$ in slits of width $D$ against scaling
variable $\psi = N^{5/2}(D+\delta)^{2(1-\nu_2/\nu)}$. Since a tethered chain
relaxes effectively as a free chain of doubled length, the vertical offset of
the straight lines in log-log coordinates is $2^{5/2}$, as expected - see text.
(b) Characteristic time of the fast chain relaxation perpendicular to the slit
wall against width $D+\delta$ for chains of different length $N$. The straight
line indicates the theoretically expected slope of $2+\nu^{-1}$.\label{tau}}
\end{figure}

All the longitudinal relaxation times evaluated from the slopes of the
gyration radius ACFs are plotted in Figure~\ref{tau} against the scaling
variable suggested by Eq.~\ref{tau scaling} The best fit for the relaxation
times of non-grafted chains is
\begin{equation}
\tau_{\|}=0.0053\psi .\label{tau_free}
\end{equation}

It is natural to expect that the effect of grafting one of the chain ends is
dynamically equivalent to doubling the chain length as far as longitudinal
relaxation is concerned. In this picture, the grafted end is similar to the
middle monomer of the doubled chain whose motion relative to the center of mass
contributes very little to the lateral size relaxation. The data points for
grafted chains are described very accurately by an equation based on this
picture,
\begin{equation}
\tau_{\|}=2^{5/2} 0.0053\psi .
\end{equation}
The global translational diffusion coefficient of the confined chain was
obtained from the mean square displacement of the center of mass as a function
of time. All the curves are nearly ideal straight lines. According to the
Einstein-Smoluchovsky equation for a $2$-dimensional diffusion
\be
   \langle( r_{cm}(t)-r_{cm}(0))^2\rangle = 4D_{\mbox{diff}}\;t
=\frac{4}{N\zeta_0}t
\label{EinSmol}
\ee
The friction coefficient in our MD simulation comes from the thermostat
coupling and is set to be equal to $0.5$. Indeed, all the simulation results are
in perfect numerical agreement with the theory. It is a commonly accepted
convention to define characteristic diffusion time as a time required to reach
the mean-square displacement equal to $R^2$. From this convention we obtain
the following scaling fit for our model:
\be
   \tau_{diff} = 0.02\psi  \label{taudiff}
\ee
Comparing  Eqs.~\ref{tau_free} and \ref{taudiff},  we find that the ratio of the
lateral diffusion time to the time of lateral relaxation amounts to a constant
factor of about $3.8$.

A scaling description for the normal relaxation is based on the idea that,as
far as trasversal motion is concerned, blobs relax independently. This would
mean that both the normal end-to-end distance and the normal gyration radius are
characterized by the same relaxation time of a single blob,
\be
   \tau_{\bot}\sim g^{1+2\nu}\sim D^{2+1/\nu}  \label{taunorm}
\ee

As mentioned above, the shape of the transversal ACFs if far from being simple
so that extracting a characteristic time requires caution. In order to evaluate
the proper normal relaxation times we were subtracting the long-time tails due
to coupling with longitudinal modes. After this subtraction, the shape of the
ACFs allows for a better defined dominant transverse relaxation time.

The transverse relaxation times are presented in Figure~\ref{tau}b against the
slit width $D$ on a log-log scale for various chain lengths. The simulation data
is consistent with the scaling prediction of Eq.~\ref{taunorm} which gives
$\tau_{\perp} \propto D^{2+1/\nu}$, being is marked by a dashed best fit line in
Fig.~\ref{tau}b: $\tau_{\perp} = 0.002 D^{2+1/\nu}$. However, the accuracy is
not very high as indicated by the error bars.

One can estimate numerically the typical time for normal relaxation, making use
of Eq.~\ref{tau_free}, by noting that $\tau_{\parallel}$ describes the
relaxation of a single blob, provided one uses in $\psi$ the number of monomers
in a blob. This yields
\begin{equation}
 \tau_{\perp} = 0.0053 D^{2+1/\nu} .
\end{equation}
which is proportional to the best fit line in Fig.~\ref{tau}b by a factor of
$\approx 2.7$.

\section{Summary and Discussion}
\label{discuss}

In the present work we present a comprehensive study of the static and dynamic
properties of flexible polymer chains confined in a narrow slit with
impenetrable repulsive walls. A combination of extensive MD simulations and an
analytic theory provides a consistent picture of polymer behavior parallel and
normal to the slit walls.
\begin{itemize}
 \item The confinement free energy (per blob) has been obtained rather
precisely and shown to be equal very nearly to $2k_BT$.
 \item Exact, model-independent expressions with no adjustable parameters are
derived for the normal components of the mean end-to-end distance and radius of
gyration, and shown to be in excellent agreement with simulation data. Also in
the case of grafted chains, closed analytic expressions for the end-to-end
distance in normal direction were obtained for arbitrary positions of the
grafting monomer.
\item The observed size of the polymer parallel to slit walls is found to
comply very well with scaling predictions in a broad interval of chain lengths
and slit widths. The corresponding probability distribution functions are found
analytically and confirmed by means of our computer experiments.
\item The bond orientation profile across the slit was predicted analytically
for Gaussian and real chains, and verified by simulation. The predicted average
orientation for different slit widths has been found to agree very well with
simulation data.
\item The characteristic relaxation times of confined chains in directions
parallel and normal to the slit walls have been obtained from evaluation of the
respective autocorrelation functions. The effect of grafting on lateral
relaxation time, which should be equivalent to doubling of the chain length, has
been demonstrated.
\item  A novel feature is the observed coupling of normal and lateral modes
with vastly different relaxation times.
\item It is found that the mean diffusion times in lateral direction, albeit
scaling similarly to the lateral relaxation time with chain length and slit
width, are larger than the latter by a constant factor of about $3.8$. The
lateral relaxation time, evaluated for a single blob, is larger by the
perpendicular relaxation time by a factor of $2.7$.
\end{itemize}

It is interesting to note that the possibility to derive accurate analytic
expressions for the static chain properties perpendicular to the slit planes is
due to the screening of correlations between distant blobs in the direction of
compression.

As mentioned before, the properties of the present system of a confined polymer
are entirely determined by the underlying anisotropy in space. The situation is
similar to that of adsorbed polymers on a plane where the thickness of the
adsorbed layer is determined by the attraction to the surface and the chain
conformations undergo deformation in perpendicular direction~\cite{AMKB2}.
Therefore it is not surprising that the adsorbed chain dynamics is described by
scaling theory in close analogy to the present treatment, as shown recently by
Descas et al.~\cite{Sommer}.

\section*{Acknowledgments}

We, L. I. K. and A. M. S.,  are grateful to the Deutsche Forschungsgemeinschaft
(DFG) for financial supportunder Grant Nos. 436 RUS 113/863/0. A. M. S. received
partial support under Grant NWO-RFBR 047.017.026. One of us (D.~D.) appreciates
support from the Max Planck Institute of Polymer Research via MPG fellowship,
another (A. M.) received partial support from the DFG under project No.
436 BUL 113/130.

\section*{Appendix}

In order to evaluate an indirect orienting effect of the walls we take into
account the fact that a bond experiences a torque due to the two tails attached
to it. This torque is non-uniform as the partition functions of the tails depend
on the position in the normal direction. The non-normalized weight for all
configurations with the $n$-th monomer at position $z$ and the $n+1$-st monomer
at position $z'=z+a \cos\theta$ is given by the product of the partition
functions of the two tails
\be
	W(z,\theta)=Q_n(z)Q_{N-n}(z+a \cos\theta)
\ee
Here we assume factorization which is exact for ideal chains. (This is also a
plausible scaling ansatz for real chains in the strong confinement limit
provided only normal directions are concerned). The partition function $Q(z)$ is
in fact the Green's function integrated over the positions of the other tail
end. Up to a normalization factor it coincides with the probability distribution
of free chain ends $\rho_{\rm end}(z)$, and is therefore a function of the form
$Q(z)=const D^{-1}f(z/D)$. For a confined ideal chain in the ground state
approximation it is independent of $n$ and is given by the ground state
wavefunction of the Edwards' equation:
\be
	Q(z)=const. D^{-1} \cos(\pi z/D)
\ee

Expanding $Q(z+a \cos\theta)$ up to second order in $\cos\theta$ we obtain
\be
W(z,\theta)=Q^2(z)\left[1+\frac{a}{D}\frac{f'}{f} \cos\theta +
\frac{1}{2} \left(\frac{a}{D}\right)^{2} \frac{f''}{f}\cos^2\theta\right]
\ee

Calculating the average $\langle \cos^2\theta\rangle$ with the weight $W$ up to
the lowest non-trivial order in $a/D$ gives the following result for the average
Legendre polynomial
\be
	\langle P_2(z)\rangle = \frac{1}{15}
\frac{f''(z/D)}{f(z/D)}\frac{a^2}{D^2}
\ee

In the general case this quantity is still a function of the $z$ coordinate and
has to be averaged over the slit width with the weight given by the full monomer
density $\rho(z)$. However, for an ideal chain, the ratio $f''/f=-\pi^2$ is
constant at any position within the slit and won't be affected by averaging.
This brings the final result
\be
	\langle P_2\rangle_{\rm id}=-\frac{\pi^2 a^2}{15 D^2} \label{pi}
\ee
which means that the indirect orientation effect is dominant compared to the
local wall effect $\sim D^{-3}$

For a real chain, after averaging over the slit width we get a result very
similar to Eq.~\ref{pi}. Only the numerical coefficient changes slightly but the
$D^{-2}$ dependence persists. Since the propagating effect of the wall is much
stronger in this case, the indirect orientation by torque constitutes just a
minor correction.

\end{document}